# Effect of Sublevel Population Mixing on the Interpretation of Doppler-Shift Spectroscopy Measurements of Neutral Beam Content.

S. Polosatkin

*Budker Institute of Nuclear Physics SB RAS*
*Novosibirsk State Technical University*
*Novosibirsk State University*
*s.v.polosatkin@inp.nsk.su*

*Interpretation of Doppler-Shift spectroscopy measurement of neutral beams species content requires an assumption about distribution of populations of thin structure of excited state (hydrogen n=3). In the paper several effects caused mixing of sublevel population are discussed and correction factors for different models of thin structure population are calculated. Such mixing can lead to tens percent uncertainty of results of measurements of beam species content. A possible way for experimental verification of sublevel population mixing is proposed.*

## Introduction

Beams of fast hydrogen atoms are widely used in nuclear fusion research for plasma heating and diagnostics. Such beams contain aside from fast atoms with full energy equal to accelerating voltage of ion source also the atoms with fractional energies (E/2, E/3, E/18) that appears in the beam due to acceleration and follows dissociation of molecular ions $H_2^+$, $H_3^+$, $H_2O^+$. Species content (that is relative concentration of such atoms) is important parameter of the beam required for determination of energy release and density of fast neutrals in plasma. Doppler-Shift spectroscopy is widely exploited for measurements of beam species content. This technique is based on measurements of intensities of spectral lines with different Doppler shift radiated by fast atoms with different velocities. Species content can be found from these measurements with the use of correction factors that have derived from effective cross-sections of excitation of observed spectral line. These correction factors depend on distribution of population of thin structure sublevels of excited state. The source of such dependence is explained in the section 2, estimation of the errors of beam species content measurements related with thin structure population model is presented in section 3, several effects that cause sublevel population mixing are discussed in section 4.

## 1. Doppler-shift spectroscopy of neutral beams

As a rule, Doppler-Shift spectroscopy diagnostic is placed after a neutralizing target and a deflecting magnet that used for separation of charged particles from the beam. In what follows we will assume that neutralizing target thick enough for complete dissociation of fast hydrogen molecules and establishing of equilibrium content of atoms and ions in the beam. For composite beam contained equilibrium amount of atoms and ions the two main processes cause spectral line radiation. That is excitation of fast atoms with following radiative decay:

**H**+H$_2$ -> **H***+H$_2$ -> **H** + H$_2$ + hν                              (1)

and radiative capture of electron by fast ion:

**H**$^+$+H$_2$ -> **H***+H$_2^+$ -> **H** + H$_2^+$ + hν                              (2)

(bold symbols mark fast particles).

Here we have neglected the processes with negative ions due to their small contribution to the intensity of line radiation. If the deflecting magnet is energized the ions have extracted from the beam and only the first process can lead to line radiation.

Actually hydrogen H$_\alpha$ line (transition n=3 -> n=2) exploits exclusively for species content measurements due to simplicity for detection of this line and availability of information about cross-sections of the line excitation. Herein in the majority of publications (e.g.[1-3]) the correction factors have taken from the work of Uhlemann ([4]), where they calculated with the use of experimental cross-section data from [].

The correction factors calculate as the ratio of effective cross-sections of H$_\alpha$ generation on the full and fractional energy normalized to fraction of particles with corresponding energies. Using the notation from the work of Uhlemann [4] and assuming full dissociation of fast molecules in the neutralizing target the correction factors can be expressed as follows:

$$c_k = \frac{s_1^0(E)\cdot f_1^0(E)+s_1^+(E)\cdot f_1^+(E)}{s_1^0(E/k)\cdot f_1^0(E/k)+s_1^+(E/k)\cdot f_1^+(E/k)}, \qquad k=1,2,3,18$$

$$c_{k0} = \frac{s_1^0(E)\cdot f_1^0(E)}{s_1^0(E/k)\cdot f_1^0(E/k)}$$

Here $c_k$, $c_{k0}$ – correction factors for composite and pure neutral beam, $\sigma_1^0$, $\sigma_1^+$ effective cross-sections of H$_\alpha$ generation by fast atoms and ions, $f_1^0$, $f_1^+$ - conversion factors on neutralizing cell, that determines as

$$f_1^0(E)=\frac{n_{H0}(E)}{h_{H+}}, f_1^0(E/2)=\frac{n_{H0}(E/2)}{h_{H2+}}, f_1^0(E/3)=\frac{n_{H0}(E/3)}{h_{H3+}}$$

$$f_1^+(E)=\frac{n_{H+}(E)}{h_{H+}}, f_1^+(E/2)=\frac{n_{H+}(E/2)}{h_{H2+}}, f_1^+(E/3)=\frac{n_{H+}(E/3)}{h_{H3+}},$$

where n – concentrations of fast particles in the point of observation, η - concentrations of corresponding particles in the ion source of injector.

The correction factors set relation between observed intensities of Doppler-shifted spectral lines and concentration of species in the ion source of injector. For measurements in pure neutral beam relative concentrations of different species can be found as

$$\frac{h(H_k^+)}{h_\Sigma} = \frac{c_{k0} \cdot I_k}{\sum_k c_{k0} \cdot I_k}$$

where $I_k$ – intensities of the corresponded spectral lines; $c_k$ instead of $c_{k0}$ should be used for measurements in composite beam that contains both atoms and ions. Notice that, according traditions of Doppler-Shift spectroscopy diagnostics, results of measurements gives as relative concentrations of beam species ($H^+$, $H_2^+$, $H_3^+$, $H_2O^+$) on the exit of the ion source, before neutralizing and dissociation.

## 2. Models of thin structure population

The aim of this work is estimation of influence of effects of re-distribution of population of thin structure sublevels to an accuracy of beam species content measurements. The level n=3 of atomic hydrogen have 5 sublevels separated by ~$10^{-5}$ eV. Parameters of these sublevels are presented in the table ([6]).

Table 1 Hydrogen n=3 level stucture

| Sublevel | Energy, eV | $g_u$ | Sublevel lifetime, ns | Sublevel branching ratio | Transition (bottom level) | Transition probability, $s^{-1}$ |
|---|---|---|---|---|---|---|
| 3s 1/2 | 12.087494193 | 2 | 158 | 1 | 2p 1/2 | 2.10e+6 |
|  |  |  |  |  | 2p 3/2 | 4.21e+6 |
| 3p 1/2 | 12.087492891 | 2 | 5.28 | 0.12 | 2s 1/2 | 2.24e+7 |
|  |  |  |  |  | 1s 1/2 | 1.67e+8 |
| 3p 3/2 | 12.087506332 | 4 | 5.28 | 0.12 | 2s 1/2 | 2.24e+7 |
|  |  |  |  |  | 1s 1/2 | 1.67e+8 |
| 3d 3/2 | 12.087506310 | 4 | 15.45 | 1 | 2p 1/2 | 5.39e+7 |
|  |  |  |  |  | 2p 3/2 | 1.08e+7 |
| 3d 5/2 | 12.087510790 | 6 | 15.45 | 1 | 2p 3/2 | 6.47e+7 |

The sublevels *3p* predominantly decay directly to ground state *1s* with emission of ultraviolet Layman-β line, which can't be detected by Doppler-Shift spectroscopy systems. Probability of radiative *3p->2s* transition (branching ratio) is equals to 0.12; *3p* sublevel excitation lifetime is 5,4 ns. Instead of *3p*, transitions from *3s* and *3d* sublevels to ground state are forbidden by selection rules, so branching ratio for these sublevels are equal to unity. Herein excitation lifetime of *3s* sublevel is equal to 158 ns, that 30 times exceed the lifetime for *3p*.

In the Uhlemann's calculations of correction factors it is assumed that there are no transitions between the level sub-structure. In this case (we will refer it as **Coronal Equilibrium model** - **CE**) cross-section of $H_\alpha$ emission can be found as follows:

$$s_{H\alpha}^{CE} = s_{3s} + s_{3d} + s_{3p} \frac{A_{3p-2s}}{A_{3p-2s} + A_{3p-1s}} = s_{3s} + 0.12 \cdot s_{3p} + s_{3d}$$

where $s_{3s}$, $s_{3p}$, $s_{3d}$ – cross-sections of excitation to corresponding sub-level, $A_{3p-2s}$, $A_{3p-1s}$ – transition probabilities for $H_\alpha$ and $L_\beta$.

Interesting feature of a coronal equilibrium is a transport of excitation away from a zone of collisions (for example, 30 keV atom excited to *3s* sublevel passes 40 cm before photon emission). According this feature often established (e.g.[7]) that for correct Doppler-Shift measurements observation point should be distanced from neutralizing target to at least several tens of centimeters. Actually, exponential decrease of $H_\alpha$ intensity with distance from gas cell was observed in several precise experiments oriented to measuring of hydrogen excitation cross-sections [8]. Moreover all experimental sublevel-resolved excitation cross-section data are derived from observation of spatial decrease of $H_\alpha$ intensity. At the same time for powerful beams several processes can cause transitions between sublevels and mixing of sublevel population. In this case (we will refer it as **Thermal Equilibrium** - **TE**) probability of a sublevel population is proportional to its statistical weight. For thermal equilibrium model cross-section of $H_\alpha$ emission can be found as:

$$s_{H\alpha}^{TE} = (s_{3s} + s_{3p} + s_{3d}) \frac{g_{3s} A_{3s-2p} + g_p A_{3p-2s} + g_{3d} A_{3d-2p}}{g_{3s} A_{3s-2p} + g_{3p} A_{3p-2s} + g_{3d} A_{3d-2p} + g_{3p} A_{3p-1s}} = 0{,}43 \cdot (s_{3s} + s_{3p} + s_{3d})$$

where g- statistical weights of corresponding sublevels.

An intermediate case that we will specify as **Collision-Radiative Equilibrium** (**CRE**) is realized if inverse lifetime of 3p level exceeds probabilities of non-radiative transitions $C_{3s-3p}$ and $C_{3d-3p}$. In the assumption of small probabilities of non-radiative quadruple transitions *3s-3d*, *3d-3s* [9] cross-section of Ha emission can be estimated as

$$s_{H\alpha}^{CRE} = s_{3s} \frac{A_{3s-2p}}{A_{3s-2p} + C_{3s-3p}} + s_{3s} \frac{C_{3s-3p}}{A_{3s-2p} + C_{3s-3p}} \cdot \frac{A_{3p-2s}}{A_{3p-2s} + A_{3p-1s}} + \dots$$

$$s_{3d} \frac{A_{3d-2p}}{A_{3d-2p} + C_{3d-3p}} + s_{3d} \frac{C_{3d-3p}}{A_{3d-2p} + C_{3d-3p}} \cdot \frac{A_{3p-2s}}{A_{3p-2s} + A_{3p-1s}} + s_{3p} \frac{A_{3p-2s}}{A_{3p-2s} + A_{3p-1s}}$$

Next if $A_{3s-2p} \ll C_{3s-3p}$ and $A_{3d-2p} \gg C_{3d-3p}$ this equation reduces to follows:

$$s_{H\alpha}^{CRE} = s_{3s} \frac{A_{3p-2s}}{A_{3p-2s} + A_{3p-1s}} + s_{3p} \frac{A_{3p-2s}}{A_{3p-2s} + A_{3p-1s}} + s_{3d} = 0.12 \cdot s_{3s} + 0.12 \cdot s_{3p} + s_{3d}$$

Actually the later means that, in contrast to coronal equilibrium, excitation to 3s level predominantly decays via non-radiative transition to *3p* with following radiation transition to *2s* or *1s*.

## 3. Correction factors calculations

Since excitation cross-sections to *3s*, *3p*, and *3d* sublevels have different energy dependence the correction factors of Doppler Shift spectroscopy measurements depend on the thin structure

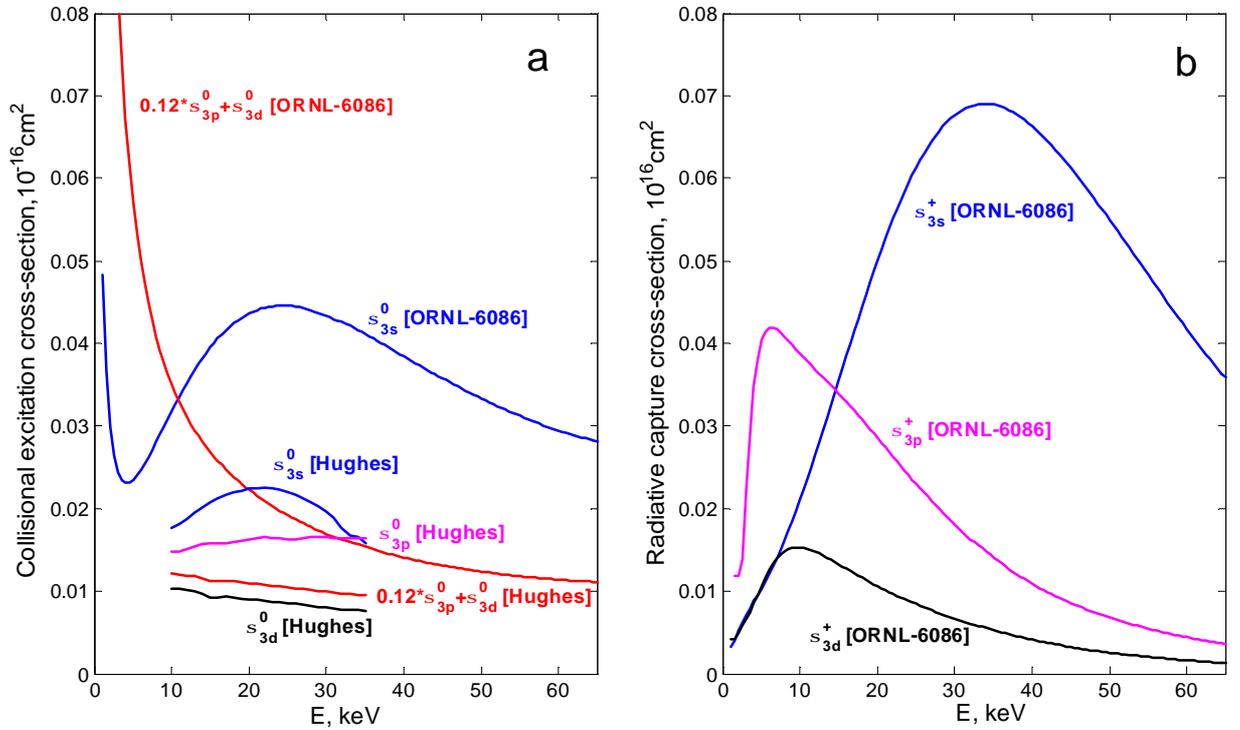

Fig.1 State-resolved cross-section of collisional excitation (a) and radiative capture (b).

population. Unfortunately there are no adequate data for sublevel-resolved collisional excitation cross-sections $s_{3s}$, $s_{3p}$, $s_{3d}$ that are required for calculation of the correction factors for different models of population. Cross-section data from [5, 10], that were used in the majority of experimental measurements (e.g.[1-4]) include only values for $s_{3s}$ and $0.12s_{3p}+s_{3d}$. The unique sublevel-resolved measurements of all $s_{3s}$, $s_{3p}$, and $s_{3d}$ cross-sections [8] covers only limited energy range 10-35 keV and disagree in absolute value with data from [5] (see fig.1). To overcome this problem we assume for our estimations that the ratio of collisional excitation cross sections $\sigma_{3p}/\sigma_{3d}$ is constant and equal to 2. This assumption based on the relations of Huges's data in the available energy range as well as conformity of *3p* and *3d* cross-sections for radiative capture. Accordingly the generalized data from [10] where used for radiative capture and collisional excitation to *3s* sublevel. Excitation cross-section to *3p* and *3d* taken as

$s_{3p}=1.6 \cdot s_{pd}$,    $s_{3d}=0.8 \cdot s_{pd}$,

where $s_{pd}=0.12 \cdot s_{3p}+s_{3d}$ also got from [10].

Calculated correction factors vs. accelerating voltage for CE and TE models were plotted on the Fig.2. For half- and third- energy atoms ($c_{20}$ and $c_{30}$ factors) discrepancy up to 30% is observed in the energy range 15-25 keV. Measurements in composite beam (without **H**$^+$ separation) are more sensitive to the sublevel population mixing. Most notable effect – up to 50% is found in the $c_{18}$ and $c_{180}$ factors that used for estimation of impurities content in the beam.

## 4. Mechanisms of population mixing

Consider possible processes that can cause the mixing of sublevel populations. These processes should have rather specific properties – large cross-section sufficiently exceeds gas-kinetic together with very small ($10^{-6} - 10^{-5}$ eV) interaction energy. Actually collision frequency $\sim 10^7$ s$^{-1}$ required to dominate of this process over radiative decay correspond to cross-section value $\sim 10^{-13}$ cm$^2$ for 50 keV beam and target particles concentration $10^{12}$ cm$^{-3}$ (that is typical concentration of residual gas in the beam drift space).

Inelastic collisions with charged particles (Coulomb centers) could satisfy these conditions. Electric field of charged particle can cause Stark shifting of sublevel energies and their overlapping. The value of electric field required for sublevel overlapping can be estimated from following condition: potential difference over the scale of excited atom should be equal to energy gap between sublevels ($10^{-6}$ eV for $3s_{1/2}$-$3p_{1/2}$ transition). That is

$$E_s = \frac{\Delta j}{n^2 \cdot 2 r_a} = 10 [V/cm]$$

That gives corresponding cross-section and concentration of Coulomb centers required for population mixing

$$\sigma = \pi r_0^2 = \pi \frac{1}{4\pi\varepsilon_0} \frac{e}{E_s} = 4 \cdot 10^{-8} cm^2 \qquad n = \frac{t}{\sigma v} \sim 10^6 cm^{-3}$$

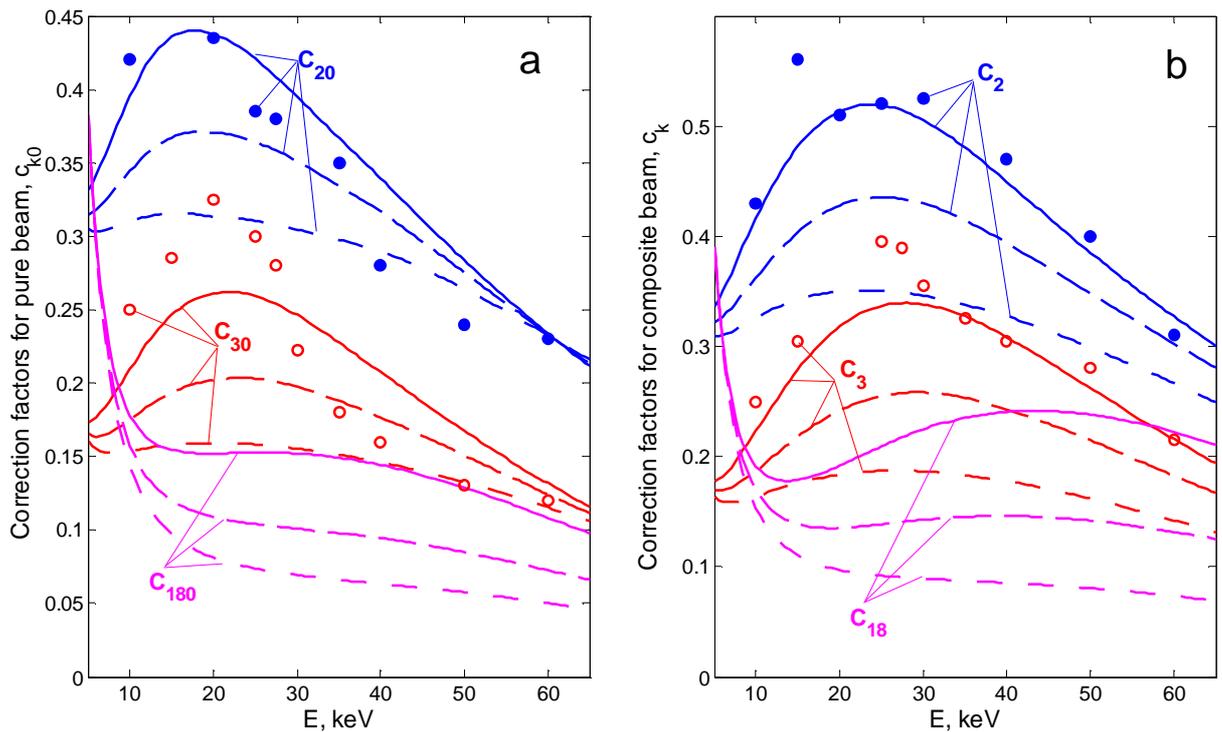

Fig.2 Correction factors for pure (a) and composite (b) beams for different models of sublevel populations; solid lines – **CE**, dashed lines – **TE**, doted lines – **CRE**; circles – correction factors from ref.???

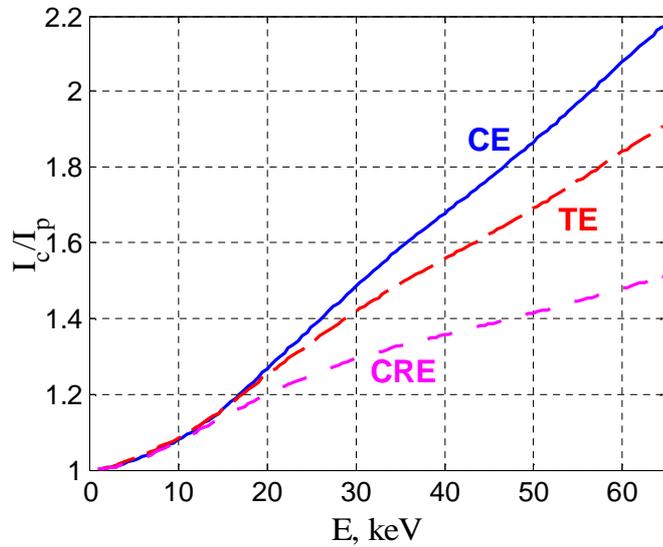

Fig.3 The ratios of intensities of Ha radiation of composite and pure neutral beam, calculated for different models of sublevel population (solid lines – **CE**, dashed lines – **TE**, doted lines – **CRE**)

Its easy to show that the density of a plasma appeared due to ionization of residual gas is much greater than the value required for sublevel population mixing.

Similar process that can cause population mixing is collisions with polar molecules. For characteristic dipole moment 1 Debye the electric field decreases to 10 V/cm in a distance $5 \cdot 10^{-6}$ cm that gives effective cross-section of population mixing $\sim 10^{-10}$ cm$^2$. Thence partial pressure of polar molecules (e.g. water) in order of $10^{-5}$ Pa is enough for mixing of population.

Another important effect is decrease of 3s sublevel lifetime in magnetic field due to motional Stark effect. MilliTesla-level stray magnetic fields could present in drift space of the injector, especially in the case of energized magnet-separator. Calculations of Foley and Levinton [11] shows that in magnetic field more than 2 mT branching ratio for 3s sublevel decrease from unity below the value of 0.2, that lead to transition from Coronal to Collisional-Radiative Equilibrium.

One possible way to observe the effects of population mixing is comparison of intensities of Doppler-shifted lines for pure and composite beams. Actually for radiative capture excitation to 3s sublevel dominates under excitation to *3p* and *3d* (fig.? of supplement) therefore intensity of H$_\alpha$ radiation produced in this process strongly depend on effective branching ratio of *3s* sublevel decay.

Ratio of intensities of H$_\alpha$ radiation for composite and pure beam can be found as

$$\frac{I_c}{I_p} = \frac{s_1^0(E) \cdot f_1^0(E) + s_1^+(E) \cdot f_1^+(E)}{s_1^0(E) \cdot f_1^0(E)}$$

The ratios calculated for different models of sublevel population are presented on the fig. Measurement of such ratio would determine which model of sublevel population is realized.

## Conclusion

Disregarding of sublevel population mixing effects can cause substantial errors in interpretation of Doppler-Shift spectroscopy measurements. These errors are most important for measurements of impurity content of the beam where they can lead to up to two-fold overestimation of impurity concentration. A presence of population mixing may be verified by comparison of ratios of line intensities for pure and composite beam with calculated values.

## Acknowledgements

This work was financially supported by the Ministry of Education and Science of the Russian Federation Government and the Integration Interdisciplinary Project of SB RAS № 104.